\documentstyle[aps,prb,epsf,floats]{revtex}

\begin{document}
\twocolumn[\hsize\textwidth\columnwidth\hsize\csname@twocolumnfalse%
\endcsname
\draft

\begin{title}
{Infinite-randomness quantum Ising critical fixed points}
\end{title}
\author{Olexei Motrunich$^1$, Siun-Chuon Mau$^1$, 
David A. Huse$^1$ and Daniel S. Fisher$^2$}
\address{$^1$Physics Department, Princeton University, 
Princeton, NJ 08544\\
$^2$Physics Department, Harvard University, Cambridge, MA 02138}
\date{\today}
\maketitle

\begin{abstract}
We examine the ground state of the random quantum Ising model in
a transverse field using a generalization of the Ma-Dasgupta-Hu
renormalization group (RG) scheme. For spatial dimensionality $d=2$,
we find that at strong randomness the RG flow for the quantum
critical point is towards an infinite-randomness fixed point,
as in one-dimension.  This is consistent with the results of
a recent quantum Monte Carlo  study by Pich, {\it et al.},
including estimates of the critical exponents from our RG
that agree well with those from the quantum Monte Carlo.
The same qualitative behavior appears to occur for three-dimensions;
we have not yet been able to determine whether or not it persists
to arbitrarily high $d$.  Some consequences of the
infinite-randomness fixed point for the quantum critical
scaling behavior are discussed.  Because frustration is irrelevant
in the infinite-randomness limit, the {\it same} fixed point
should govern both ferromagnetic and spin-glass quantum
critical points.  This RG maps the random quantum Ising model
with strong disorder onto a novel type of percolation/aggregation
process.
\end{abstract}

\vskip 0.3 truein
]

\section{Introduction}

Systems with quenched randomness and many degrees of freedom
may be divided into three classes based on their coarse-grained
behavior in the low-energy, low-frequency and/or
long-distance limit.  First, there are many systems where the
quenched disorder is irrelevant in the renormalization-group
sense.  Such  systems, even though they are spatially inhomogeneous
at the microscopic scale, become asymptotically homogeneous at
macroscopic scales; their coarse-grained, low-energy behavior is
the same as some ``pure'' system without quenched disorder.
In this case, the renormalization-group fixed point governing the
coarse-grained system is at zero quenched randomness.
The second possibility is systems controlled by fixed points
with nonzero, but finite, quenched randomness. In this case the
coarse-grained behavior is spatially inhomogeneous, but the
relative magnitude of the inhomogeneities remains finite at 
the fixed point. Examples of this second class include
spin glasses and other glassy phases, as well as various
critical points with randomness.
The third possibility, which is the subject of this paper, occurs
when the quenched randomness and thus the relative magnitude of the
inhomogeneities grows without limit as the system is coarse-grained.
So far, we know of a few infinite-randomness fixed points
which comprise  this third class; all of them are one-dimensional 
($d=1$) quantum ground
states. These are the random singlet states of certain random 
antiferromagnetic spin chains, \cite{mdh,f94}
the quantum critical point of the random quantum Ising (and Potts)
chain, \cite{mw,sm,dsf,pot} and quantum critical points
separating random singlet states and the Ising antiferromagnetic
phase \cite{f94} or the Haldane state in the random
spin-1 Heisenberg chain. \cite{hy,mgj}

A natural question is whether such infinite-randomness fixed
points can govern the behavior of physical systems with spatial
dimensionality $d \geq 2$. 

Here we study the simplest random quantum system that can 
exhibit a phase transition, the random  ferromagnetic 
Ising model in a transverse field, which, in all dimensions, 
has a quantum critical point at zero temperature.  We
focus on the strong-randomness regime, using a generalization 
\cite{dsf} of the Ma-Dasgupta-Hu \cite{mdh} renormalization 
group (RG) scheme.  This approximate RG is exact in the limit 
of infinite  randomness and thus can in principle yield exact 
results for the scaling behavior of systems governed by infinite 
randomness fixed points.\cite{dsf}  In one dimension,
this RG has been used to analyze the various infinite-randomness 
fixed points mentioned above and many results can be obtained 
analytically, in particular for the random quantum Ising model. 
In higher dimensions, the renormalization group   
cannot (to our knowledge) be carried out analytically; in this 
paper we analyze its general structure and consequences, 
implement it numerically and examine a simple  
approximation to it.  For $d = 2$ and 3, we find that the 
renormalization group flow on the critical manifold for strong
randomness  is indeed towards even stronger randomness, as in
one-dimension, indicating that the quantum critical behavior 
is governed by the infinite-randomness critical fixed point. 
For $d=2$ we have studied the RG flow thoroughly enough to be 
fully confident that this is the case, and it appears to remain 
true for $d=3$. 

For $d < 4$, the Harris criterion \cite{har} indicates that
at the pure Ising quantum
critical point weak randomness is relevant, with the
RG flow towards stronger randomness; thus the simplest scenario for
two- and three-dimensions is that the infinite-randomness fixed
point governs the random quantum critical point with {\it any}
randomness, as in one-dimension.
A recent quantum Monte Carlo study by Pich,~{\it et~al.}
of the ferromagnetic model with moderate randomness in
two-dimensions is consistent with this picture. \cite{py}

What do we mean by infinite randomness?  This means
\cite{dsf,f94} that as the system is coarse-grained and
the characteristic energy scale decreases, the distributions of
the {\it logarithms} of the magnitudes of the terms in the 
renormalized hamiltonian become arbitrarily broad.  As a result,
the ratio of the magnitude of any two terms approaches either 
zero or infinity. In this limit each renormalized
coupling completely dominates any weaker renormalized coupling, 
so even the Ising spin glass becomes unfrustrated.\cite{ns}  
Thus we expect that the same infinite-randomness fixed point 
governs both  the random ferromagnetic and the spin-glass quantum 
critical points, with simple modifications to account for the
antiferromagnetic bonds in the latter case.
We should note that recent quantum Monte Carlo studies of the 
spin-glass case had concluded that for two and three dimensions
the scaling near the quantum critical point is 
conventional,\cite{qsg} implying a finite-randomness
fixed point, in contrast to what we are proposing here.
But these finite-size scaling Monte Carlo studies looked at
a rather small size range and did not look at distributions of 
physical properties; thus they were not sensitive to the scaling 
towards infinite randomness
that we now believe occurs for $d=2$ and probably for higher $d$.

\section{Cluster RG }   

We will study the quantum spin-1/2 Ising model with random 
ferromagnetic interactions, positive transverse fields $h_i$, 
and moments $\mu_i$:

\begin{equation}
H = -\sum_{i<j}J_{ij}\sigma_i^z\sigma_j^z - \sum_ih_i\sigma_i^x - 
H\sum_i\mu_i\sigma_i^z  ~,
\end{equation}
where the $\sigma$'s are the Pauli spin matrices.
The interactions $J_{ij}$ dictate the geometry and thus the
dimensionality of the system, with the strong interactions being
predominantly between nearby pairs of spins. The uniform ordering
field $H=H_z$ is zero or small; it is included only to probe
the system's magnetization and susceptibilities.

The cluster RG finds the system's ground state by 
successively eliminating the highest energy
degrees of freedom. At each step, we find the largest term in the
hamiltonian, which is either a transverse field or an interaction; 
its strength $\Omega$  sets the (maximum remaining) energy scale. 
If the largest term is the field on spin $i$, $\Omega=h_i$, 
that spin is put in the ground state, $\sigma_i^x = 1$, of the 
local field term and virtual excitations to the other
state ($\sigma_i^x = -1$) are treated in second-order 
perturbation theory. For this to be a valid approximation, 
the field must be much stronger than all this
spin's interactions, which is true only in the strong-randomness 
limit. This step eliminates the spin $i$, and generates new 
interactions of the form $J_{jk}'\approx J_{ji}J_{ik}/h_i$.
But some of these new interactions may be negligible compared to 
the interactions that were already present, so the
full renormalized interactions are given, for each pair $(j,k)$, by

\begin{equation}
\label{newJ_sitedec}
J_{jk}' \approx {\rm max} (J_{jk},{J_{ji}J_{ik} \over h_i}). 
\end{equation}
We use the maximum here because in the strong-randomness 
limit the sum of two nonnegative numbers of very different
magnitudes is  well approximated by the larger number.

If, on the other hand, the largest term in the hamiltonian
is an interaction, $\Omega=J_{ij}$, the two spins involved
are combined into one new spin---a cluster---whose two states
represent the two ground states ($\sigma_i^z=\sigma_j^z=\pm 1 $)
of that interaction.  Again, the virtual excitations to the
states that are eliminated are treated in second-order
perturbation theory. With the new spin labeled $i$ 
(an arbitrary choice) the renormalized field is 

\begin{equation}
\label{newh}
h_{i}' \approx {h_ih_j \over J_{ij}} ~,
\end{equation}
and the renormalized moment is simply
 
\begin{equation}
\mu_{i}'=\mu_i+\mu_j.
\end{equation}
The renormalized interactions are, for each remaining $k$,

\begin{equation}
\label{newJ_bonddec}
J_{ik}'\approx\max(J_{ik},J_{jk}).
\end{equation}

The net result in both cases is the elimination of one spin along 
with the various renormalizations and reconnections of the lattice.
The major complication for $d > 1$ is that the RG does
not preserve the lattice structure but instead generates a disordered
and strongly correlated network of sites---i.e. clusters---and bonds
connecting them.

The action of the cluster RG is a novel aggregation/annihilation
process.  When the strongest term is a field, the corresponding
cluster is removed (annihilated), while when it is an interaction
the two clusters that it connects are aggregated into one cluster.
The clusters thus represent sets of the original spins that are
strongly correlated.  In the paramagnetic phase, the annihilation
process dominates and no large clusters are formed,
while in the ordered phase the aggregation dominates
producing arbitrarily large clusters.  For $d > 1$,
in the ordered phase an infinite percolating cluster
appears at a {\it finite} energy scale.  When this occurs,
the infinite cluster is represented by a single renormalized
spin that has an infinite coordination number and an
infinite magnetic moment.  Thus the topology of the network has by
this energy scale completely changed from that of the ``bare''
$d$-dimensional lattice which had only short-range interactions.
This emphasizes that this RG is not simply a ``real-space'' RG;
it is more precisely an ``energy-space'' RG.

The zero-temperature quantum critical point is a new type of 
percolation transition at which an infinite cluster first appears
(at the quantum critical point it only appears in the zero energy 
limit), as pointed out by Monthus, {\it et al.}\cite{mgj}.
The quantum critical point occurs when the annihilation and 
aggregation processes balance, so that arbitrarily
large clusters are produced but no single cluster dominates
at any finite energy scale.

Because of the multiplicative structure of the RG recursion 
relations, it is convenient to write the renormalized 
interactions after decimating down to energy scale $\Omega$ in 
logarithmic variables as: 

\begin{eqnarray}
h_i=\Omega e^{-\beta_i}, \\
J_{ij} =\Omega e^{-\zeta_{ij}} .
\end{eqnarray}  

The RG equations involve the full joint distribution 
of the positive quantities $\{\beta_i\}$ and $\{\zeta_{ij}\}$.
The single-field distribution function $R(\beta)$ and the 
single-bond distribution function $P(\zeta)$ provide partial but
important information about the full joint distribution of all 
the couplings. In general, the coordination number is not fixed, 
and if there are $N$ spins (clusters) remaining, there may be up 
to $N(N-1)/2$ interactions.  However, only of order $N$
of the strongest interactions matter because the weaker ones will
be overruled by the stronger ones in the action 
of the RG for strong randomness.
Thus we normalize the bond distribution function $P(\zeta)$
{\it per remaining spin}, so that the total number of bonds in 
the system is $N \int_0^\infty P(\zeta) d\zeta$, and 
for $d>1$ we define
the ``width'', $w$, of this bond distribution by 

\begin{equation}
\int_0^w P(\zeta)d\zeta = 1.
\end{equation}
This width thus includes only the $N$ strongest bonds.
[Note that in one dimension, there are always exactly $N$ 
nearest neighbor bonds and this normalization of $P(\zeta)$ 
coincides with that of Ref.\cite{f94}.  If further neighbor bonds 
exist in one dimension, they will be irrelevant at low energy 
scales and can be ignored.] 
The condition for validity of our approximate RG is that
the widths of the distributions $R(\beta)$ and $P(\zeta)$ be
large; for the RG to be asymptotically exact, these widths
should tend to infinity as $\Omega\to 0$; this is the indication 
that the RG flow goes to an infinite-randomness fixed point.
We now discuss the general structure of such a putative
infinite-randomness critical fixed point.
We will assume that the simplest scaling scenario for an
infinite-randomness fixed point occurs in our cluster RG
\cite{stph}---this is the case in one-dimension,\cite{dsf} and is 
consistent with our numerics for two-dimensions.

\section{Critical Fixed Point}

We first consider the behavior at the quantum critical point. 
We expect that at the critical fixed point the distributions
$R(\beta;\Gamma)$ and $P(\zeta;\Gamma)$ will asymptotically
be given by the simple scaling forms

\begin{eqnarray}
R(\beta;\Gamma) d\beta = B(\beta/\Gamma)d\beta/\Gamma ~,\\
P(\zeta;\Gamma) d\zeta = Z(\zeta/\Gamma)d\zeta/\Gamma
\end{eqnarray}
for large $\Gamma$; where

\begin{equation}
\Gamma \equiv \ln \left(\frac{\Omega_0}{\Omega}\right) > 0
\end{equation}
is the logarithm of the energy scale, relative to an 
energy scale $\Omega_0$ set by the properties of the bare 
hamiltonian. As $\Gamma$ is increased by $d\Gamma$, the fractional
decrease in the number of spins or clusters is $(Z_0 +B_0)
d\Gamma/\Gamma$ with 

\begin{equation}
B_0\equiv B(0) , \ \ \ \ \ 
Z_0 \equiv Z(0) .
\end{equation}  
The density of remaining clusters per unit volume thus decreases
under renormalization as 

\begin{equation}
n_\Gamma \sim \Gamma^{-(Z_0+B_0)} .
\end{equation}
This gives the basic relationship between the length scale, $L$,
and the energy scale at the quantum critical point:

\begin{equation}
\Gamma=\ln\left(\frac{\Omega_0}{\Omega}\right)\sim L^\psi ~,
\end{equation}
with $\psi=d/(Z_0+B_0)<1$. 
Note that this is very different from conventional power-law 
scaling; here it is the {\it logarithm} of the energy scale 
that varies as a power of the length scale. Since this is 
associated with the ``tunneling" events by which clusters flip, 
it has been dubbed ``tunneling dynamic scaling".\cite{stph}

The typical moment of a cluster---the number of strongly correlated
``active'' spins in it (i.e., those not yet decimated)---scales as 

\begin{equation}
\mu \sim \Gamma^\phi,
\end{equation}
so that the fractal dimension of the set of active spins 
in a cluster is 

\begin{equation}
d_f=\phi\psi.
\end{equation}
This determines the decay of the average spin-spin correlation 
function at the critical point.

\subsection{Critical Correlations}

The correlation function between two spins at distance $r$,

\begin{equation} 
G_{ij} \equiv <\sigma_i^z \sigma_j^z>
\end{equation}
is a random quantity with a very broad probability distribution 
for large $r\equiv |{\bf r}_i-{\bf r}_j|$. We first consider the 
correlation function of a  {\it typical} pair of spins with 
separation $r$; typical spin pairs are never active in the
same cluster and have only weak correlations that fall off, 
at criticality, as a stretched exponential function of distance,

\begin{equation} 
\label{typ-corr}
-\ln G_{\mathrm typ}(r) \sim r^\psi,
\end{equation}
with $\psi < 1$.
These correlations arise from the lowest order perturbative 
corrections to the decimation of spin clusters \cite{dsf,fyoung}.  
When a cluster is decimated at energy scale $\Omega$, 
each of the effective spins on its neighboring clusters---in 
more conventional terms the perturbatively modified 
wavefunctions that are labeled by 
the remaining effective spins---will acquire a component of the 
decimated cluster's spin whose magnitude is of order $J/\Omega$, 
with $J$ the effective coupling that links the neighboring cluster 
to the decimated cluster. 
Likewise, when these neighboring clusters are decimated, an even 
smaller component of the original spin will be acquired by the 
remaining clusters.  Correlations between two spins $i$ and $j$ 
which are never active in the same cluster thus occur when a 
surviving cluster contains simultaneously a component of 
{\it both} the spins $i$ and $j$.  The correlation function
$G_{ij}$ is then determined by the maximum over all such mutual 
clusters that occur, at any energy scale, of the product of the 
the two spins' components contained by the mutual cluster.  
Typically, the smallest of the multiplied perturbative factors 
that determine $\ln G$ will dominate; these arise 
from the stage at which the two spins first have a component on 
a mutual cluster.  Since this occurs when the remaining cluster 
sizes are of order the separation $r$ between the spins of 
interest, the dominant perturbative factor will be of order 
$e^{-K_{ij}\Gamma}$ with $\Gamma\sim r^\psi$ and $K_{ij}$ random 
and of order one.  This yields the result 
Eq.(\ref{typ-corr}).\cite{dsf,fyoung} 

We should note that there is another mechanism by which spins 
become correlated. At any step of the renormalization, when 
a cluster with field $\tilde{h}$ is decimated, higher order 
perturbative effects will give components of the decimated spins 
on {\it all} the remaining clusters, with magnitudes that involve 
products over all the bonds connecting the decimated cluster to 
the remaining clusters, of factors of 
the form $\tilde{J}_{jk}/\tilde{h}$. 
These by themselves would give rise to simple exponential 
decay of typical correlations as occurs in conventional 
disordered phases.  In contrast to these, the contributions to the 
correlations that come from the cumulative effects of the 
successive lowest order perturbative terms discussed above 
will have similar form but with each of the $\{\tilde{J}_{jk}\}$ 
being divided by an effective field from one of the {\it later} 
stages of the RG; these are always smaller than $\tilde{h}$.  Thus 
the cumulative lowest order contributions discussed above will 
always dominate over the simple exponential decays from the 
higher order perturbative effects.  We hence conclude 
that the $\psi$ from the RG must be less than one.  

The {\it average} correlation function $\overline{G(r)}$ behaves 
quite differently than typical correlations. It is dominated by 
the rare spin pairs which are active in the {\it same} cluster 
at {\it some} energy scale; such pairs of spins have correlations 
of order one, reduced from one only by the short scale high 
energy fluctuations which are not included in the approximate RG.
As a result, the average correlation function is proportional to
the probability of the two spins being active in the
same cluster at {\it some} energy scale. At criticality this 
occurs---if at all---at scale $\Gamma\sim r^\psi$,
and the resulting average correlation function hence falls off 
as a power law:

\begin{equation} 
\label{avcorr}
\overline{G(r)} \sim r^{-\eta}\sim r^{-2(d-\phi\psi)}.
\end{equation}
This is an example of the radically different scaling behavior
of the typical and average quantities that is one
hallmark of infinite-randomness fixed points.\cite{dsf,stph}

Thermodynamic properties involve averaging over the whole system 
and will hence be dominated, as are the average correlations, 
by rare clusters.  The low temperature susceptibility to a small 
ordering field $H$ ($H \ll T$) can be found easily by stopping 
the RG at energy scale $\Omega \sim T$.  For small $T$, almost all 
the decimated spins are frozen and hence non-magnetic, while almost 
all  the remaining clusters have effective transverse fields and 
interactions between them which are much less than $T$.  They are 
hence essentially free and have independent Curie susceptibilities 
yielding an overall susceptibility at low temperatures near the 
quantum critical point of

\begin{equation}
\chi \sim \frac{|\ln T|^{2\phi-d/\psi}}{T}.
\end{equation}

The magnetization in a small ordering field $H$ (at $T \ll H$)
can be found similarly: the RG is stopped when the typical magnetic 
energy, $H\mu$, of a cluster is of order $\Omega$.  The 
decimated spins are non-magnetic while the remaining clusters 
are almost perfectly polarized by the field.
This yields, at the quantum critical point, a magnetization 
proportional to the fraction of spins which are still active,

\begin{equation} \label{McH}
M\sim  |\ln H|^{\phi-d/\psi} = \frac{1}{ |\ln H|^{\eta/2\psi} }.
\end{equation}

\subsection{Simple Approximate RG}

In order to get a better feeling for the scaling behavior, 
it is useful to study a simple approximation to the RG flows 
which is exact in one-dimension and in some respects good for 
higher $d$; this consists of ignoring correlations among 
the fields and between the fields and the bonds, but allowing
correlations among the bonds.  In this approximation, 
the evolution equation for the field distribution 
$R(\beta;\Gamma)$ depends on the bond distribution
only through $P(0;\Gamma)$, and is  identical to its 
one-dimensional form (see Refs.~\cite{dsf,stph}). Solving for 
the fixed point gives scaling distributions with $Z_0=1$ and
$B(\theta)=B_0e^{-B_0\theta}$ with $B_0$ undetermined 
(the exact solution for one-dimension has $B_0=1$).  
Within this approximation the exponents are
$\psi=d/(1+B_0)$ and $\phi=(1+\sqrt{1+4B_0})/2$.  Our numerical
studies of the RG flows in two-dimensions show that the log-field
distribution is very close to the simple exponential form, but 
the estimated critical exponents for two-dimensions differ 
somewhat from this simple approximation; this must be due to 
correlations among the fields and between them and the bonds.

\section{Numerical RG Study of the Critical Fixed Point in 
Two-dimensions}

In order to do better and certainly to test  the conjecture 
of a controlling infinite-randomness fixed point, we must 
implement the strong-randomness cluster RG numerically.  
The formulation is the same for a general random network, 
but we are of course interested in systems that can arise 
from finite-dimensional lattices with short range interactions.  
We have thus  studied  the RG flows with ``initial conditions'' 
of finite $d$-dimensional lattices for two and three dimensions.
The program has been tested by verifying that it reproduces (within
statistical errors) the analytical results for one-dimension.

For $d > 1$, many weak interactions are generated that, for
the larger lattices studied, can not all be stored. 
Because of this, we keep only interactions above a minimum 
strength $J_{\mathrm min}$, the smallness of $J_{\mathrm min}$ 
being limited by computer memory capacity and speed.  Since the 
discarded bonds (those with $J<J_{\mathrm min}$) could 
not have generated stronger bonds, for the renormalization 
down to any energy scale with $\Omega>J_{\mathrm min}$
the RG decimation sequence is not affected at all, and all
the bonds with $J_{\mathrm min} \leq J \leq \Omega$ are retained; 
thus we know all the fields and all the bonds with
$0 \leq \zeta \leq \zeta_{\mathrm m}=\ln(\Omega/J_{\mathrm min})$. 
However, under renormalization $\zeta_{\mathrm m}$ 
decreases and $\Gamma \equiv \ln(\Omega_0/\Omega)$ increases 
so that the range of $\zeta/\Gamma$ in the scaled distribution 
$Z(\zeta/\Gamma)$ that 
we can study steadily decreases as the system is coarse-grained.
We have been quite conservative and do not look at all at the
``contaminated'' low-energy part of the bond distribution 
($J < J_{\mathrm min}$).
The limits on memory are most restrictive at the earliest
stages of the RG decimation, where the number of clusters is large,
so this is when $J_{\mathrm min}$ must be set the largest.  It 
might be possible to let $J_{\mathrm min}$ decrease later in the 
decimation and recover more of the renormalized bond 
distribution with controllable errors.  We have not
explored this possibility.  

We start with initial conditions of systems of up to $10^5$
spins with random short-range interactions
and random transverse fields independently chosen from specified
initial probability distributions.  We run up to 1000 samples for
each initial probability distribution to reduce the statistical
errors.  For each sample we measure properties of the system when
the energy scale passes (under renormalization) a predefined set 
of energies \cite{fyoung}, and then average these properties 
over different samples.

To reduce transients as much as possible, the shapes of the 
initial distributions $R(\beta)$ and $P(\zeta)$ are chosen to 
approximate, as best as we can, the renormalized critical 
point distributions which we observe.  However, these initial 
conditions are missing any correlations among the fields and 
interactions which certainly exist in the full joint 
distribution at the critical fixed point.  
Thus when we run the RG it does show a fairly
strong transient behavior as these correlations are generated
and the fixed point is approached.
So far, we have only a limited understanding of these transients
and the correlations that are generated and we do not
have a systematic way of controlling them; we do, however, monitor
the simplest types of correlations and they do appear to
stabilize after the initial transient in the RG.

In our numerics we primarily
concentrate on the individual field and bond distributions 
$R(\beta;\Gamma)$ and $P(\zeta;\Gamma)$; 
these are partial but significant indicators of what is happening
in the system's full joint probability distribution. 
For two-dimensions we find that at the critical point
both distributions do become broader under the action of the RG
and the flow towards stronger randomness is clear.
This flow is weaker but nevertheless is clearly apparent for 
three-dimensions also. 
 
\begin{figure}
\epsfxsize=\columnwidth
\centerline{\epsffile{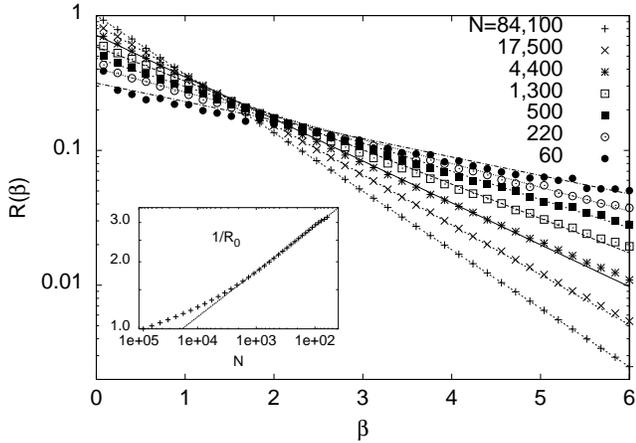}}
\vspace{0.15in}
\caption{RG evolution of the field distribution at a 
putative quantum critical point. Initial conditions are a 
$290 \times 290\ (N =84,100)$ triangular lattice with 
couplings drawn independently from the field distribution 
$R(\beta)=e^{-\beta}$ and the scaled bond distribution 
$P_{\rm sc}(\zeta_{\rm sc})=0.1+0.105 \zeta_{\rm sc}$, as
described in the text.  Lines are fits to the simple exponential 
form $R=R_0 e^{-R_0\beta}$ with $R_0$ depending on the energy scale.
Note that the renormalized 
field distribution fits this form well for all $N$. 
Inset: width of the field distribution, $1/R_0$, vs. the number 
of remaining spins $N$; the RG evolution is in the 
direction of {\it decreasing} $N$.  The increasing width indicates
the RG flow towards infinite randomness.
The line here is a power-law fit that gives our estimate
of the exponent $\psi$.  Note that this fit works well only 
after the rather strong initial transient.}
\vspace{0.0in}
\label{field_distflow}
\end{figure}

Numerically we find that under the RG the field distribution
maintains fairly accurately a simple exponential form,

\begin{equation}
R(\beta;\Gamma) \cong R_0(\Gamma) e^{-R_0(\Gamma)\beta}\;,
\end{equation}
with 

\begin{equation}
R_0(\Gamma)\equiv R(\beta=0;\Gamma).
\end{equation}
The width of the distribution is proportional to $1/R_0(\Gamma)$ 
and grows steadily as the energy scale is decreased. 
This is shown in Fig.~\ref{field_distflow} for a flow near the
critical point, but it is also true away from the critical point,
and is consistent with the simple approximation to the RG flows
discussed above for which the field distribution at low energy 
scales is always an exponential whose width never 
decreases.  

\narrowtext

\begin{figure}
\epsfxsize=\columnwidth
\centerline{\epsffile{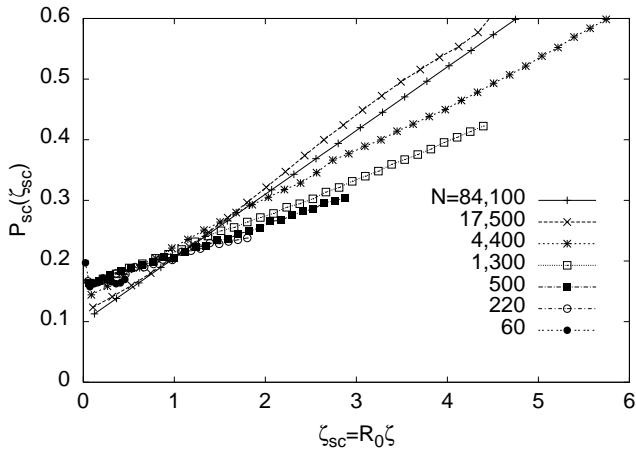}}
\vspace{0.15in}
\caption{RG evolution of the scaled bond distribution 
$P_{\mathrm sc}(\zeta_{\mathrm sc})$ at the apparent 
critical point corresponding to the initial conditions 
of Fig.~\ref{field_distflow}. Note that although the   
intercept is fairly stable at $P_{\mathrm sc}(0) \frac{\simeq}{}
 0.15$ in the
later stages of the RG (this is our criterion for locating the 
critical point---see also Fig.~\ref{intercept}), the shape of the
distribution is not as stable (see also Fig.~\ref{slope}), 
presumably due to transient effects from our uncorrelated initial 
conditions.  
Because we do not keep bonds weaker than $J_{\mathrm min}$ (see
text), the range of the scaled bond distribution that we 
measure and plot, 
$0 \leq \zeta_{\mathrm sc} \leq \zeta_{\mathrm m} = 
R_0 \log(\Omega/J_{\mathrm min})$,
decreases as we run our RG.
}
\label{bond_distflow}
\vspace{0.0in}
\end{figure}

As initial conditions, we choose for convenience the simple 
exponential distribution of log-fields with the initial $R_0=1$. 
[Note that the strong-randomness RG equations, 
Eqs.(\ref{newJ_sitedec} -- \ref{newJ_bonddec}), are invariant 
under a multiplication of all the log-couplings by any constant.  
Thus, although the initial choice $R_0=1$ would appear to 
correspond to moderate randomness, we can, without loss of 
generality, use this in our study of the strong-randomness 
RG flows.] To search for a fixed point we measure  
$1/R_0(\Gamma)$ and scale both $\beta$ and $\zeta$ 
by this width, defining

\begin{eqnarray}
\beta_{\rm sc}=R_0\beta\ \ \ \ \ \  
 {\rm and}\ \ \ \ \ \ 
\zeta_{\rm sc} =R_0\zeta.
\end{eqnarray}
(Note that $\Omega_0$ and thus $\Gamma$ are not defined precisely, 
so we can not simply scale the data by $\Gamma$.)
The scaled field distribution is now
$R_{\rm sc}(\beta_{\rm sc}) \cong \exp (-\beta_{\rm sc})$
and we can concentrate on the scaled bond distribution
$P_{\rm sc}(\zeta_{\rm sc})$.
The shape of the bond distribution evolves
continuously and its characterization is much less clear.
Numerically we observe that for $d \geq 2$ at and near the quantum
critical point the cluster RG always generates positively sloping
($dP/d\zeta > 0$) bond distributions.  This is in contrast to 
one-dimension, where the exact critical fixed point bond 
distribution 
$P_{\rm sc}(\zeta_{\rm sc}) = \exp(-\zeta_{\rm sc})$ is 
the same simple exponential as the field distribution due to a
duality relation.\cite{dsf}

For $d=2$ the bond distributions that are generated by the RG
near the quantum critical point can be reasonably approximated, 
in the small $\zeta_{\rm sc}$ regime of interest,
by a simple linear fit: 
$P_{\rm sc}(\zeta_{\rm sc}) \cong a + b\zeta_{\rm sc}$.  
We thus choose for initial conditions a $P_{\rm sc}(\zeta_{\rm sc})$ 
of this form.  Our initial 
lattice for all the data presented here is a triangular lattice 
with periodic boundary conditions
(we also tried others, such as square, to confirm that the 
results did not depend strongly on this arbitrary choice).
Since we expect the stronger bonds to be shorter-ranged,
we select the nearest-neighbor bonds (there are three such bonds
per site) to constitute the strongest-bond part 
$0 < \zeta_{\rm sc} < \zeta_{\rm c}$ of the bond distribution,
with $\zeta_{\rm c}$ chosen so that 
$\int_0^{\zeta_{\rm c}}(a+b\zeta)d\zeta=3$, i.e. there are  
precisely three bonds per site with $\zeta_{\rm sc}$ between 
zero and $\zeta{\rm_c}$.
Then the next batch of the distribution,
$\zeta_{\rm c} < \zeta_{\rm sc} < \zeta_{\rm m}$, are assigned 
at random to all the second- and third-neighbor bonds (six more bonds
per site), with $\zeta_{\rm m}$ chosen appropriately. 
(This $\zeta_{\rm m}$ sets our $J_{\rm min}$, as discussed
above.)  Thus our initial condition has nine bonds per site, 
corresponding to a coordination number of eighteen.  
Under renormalization, the lattice is quickly randomized,
so it no longer resembles the initial triangular lattice, 
and the number of bonds kept per remaining cluster initially 
increases.  
However, since we do not keep bonds with $J<J_{\rm min}$, 
the number of bonds kept per cluster decreases in the later stages 
of the decimation, as $\Omega$ decreases towards $J_{\rm min}$.
Note that our choice of the part of the distribution with 
$\zeta_{\rm sc} > \zeta_{\rm c}$ (the tail)
is only a matter of convenience; its details (and even
its very presence) are not important: 
specifically, we have checked that by the time the RG reaches this 
energy range, most of the original bonds from the tail are
gone, having been 
dominated by the stronger interactions that arise from  
the original strong first-neighbor bonds only. 

\begin{figure}
\epsfxsize=\columnwidth
\centerline{\epsffile{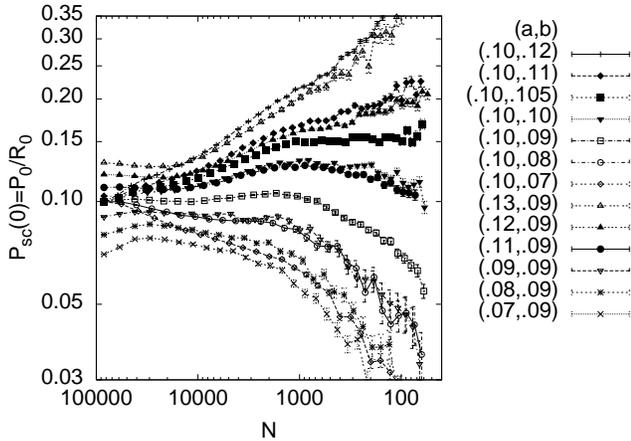}}
\vspace{0.15in}
\caption{RG evolution of the intercept $P_{\mathrm sc}(0)$ 
of the scaled bond distribution for different initial conditions: 
all are initially $290\times 290$ triangular lattices 
with independent couplings and initial field distribution
$R(\beta)=e^{-\beta}$; 
the different curves correspond to 
various initial scaled bond distributions: 
$P_{\rm sc}(\zeta_{\rm sc})=a+b\zeta_{\rm sc}$
with the parameters $a$ and $b$ as indicated (see text).
If the intercept saturates to a finite nonzero value as
$N$ is decreased, this indicates that the system is critical.
Our best estimate of the critical point has $a=0.10$, $b=0.105$
(filled squares); this is what is used in all the
other figures.  Other parameters which we view as possibly critical
are also indicated by filled symbols. The error estimates on the 
various critical exponents include the results from all of 
these potentially critical systems.}
\label{intercept}
\vspace{0.0in}
\end{figure}

\begin{figure}
\epsfxsize=\columnwidth
\centerline{\epsffile{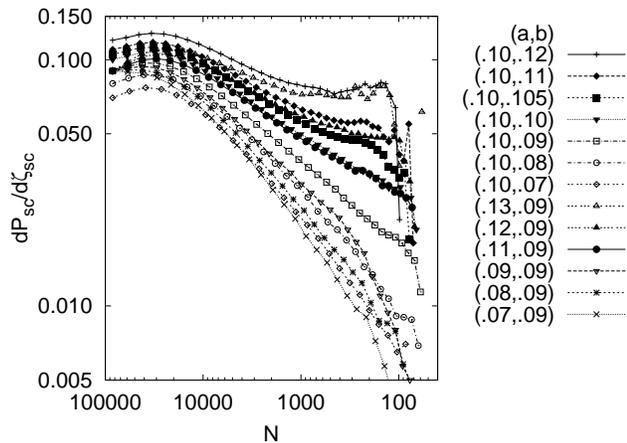}}
\vspace{0.15in}
\caption{Evolution of the slope $dP_{\mathrm sc}/d\zeta_{\mathrm sc}$
of the scaled bond distribution for the same set of different 
initial conditions as in Fig.~\ref{intercept}. At each energy scale,
the slope is calculated by fitting a linear function to 
the corresponding distribution (as in Fig.~\ref{bond_distflow}) 
in the full available region
$0 \leq \zeta_{\mathrm sc} \leq \zeta_{\mathrm m}$. 
The observed shape of the bond distribution is only approximately
linear, and our data for the bond distributions becomes very 
limited and noisy for small $N$.  The strongly transient behavior
seen here is presumably due to both actual transients in the
shapes of the distributions and the reduction with decreasing $N$
of the range, $0$ to $\zeta_{\mathrm m}$, over which the 
linear fit is made.  
}
\label{slope}
\vspace{0.0in}
\end{figure}

We first searched for a fixed point of the RG by 
starting from such a linear distribution of initially 
uncorrelated bonds and monitoring the flow
of the two parameters obtained by fitting the renormalized and
scaled bond distribution to such a linear form---the
intercept $P_{\rm sc}(0)$ and the slope
$dP_{\rm sc}/d\zeta_{\rm sc}$.
By choosing an initial bond distribution
close to the fixed point distribution we tried to
minimize the transients that occur as the full fixed point 
{\it joint} distribution of the fields and bonds is generated 
by the RG.  However, the transients remained too strong for 
us to accurately locate a fixed point of the RG flow in the 
plane of these two parameters: we could not fully 
stabilize this scaled bond distribution.  
Fig.~\ref{bond_distflow} shows an example of the 
evolution of the scaled bond distribution for initial conditions
near what we estimate to be the critical fixed point.
The intercept stabilizes at $P_{\rm sc}(0) \cong 0.15$ (see
Fig.~\ref{intercept}), but the slope is much less stable,
although it may be approaching a limit, as shown in 
Fig.~\ref{slope}.

\narrowtext

\narrowtext

\narrowtext
\begin{figure}
\epsfxsize=\columnwidth
\centerline{\epsffile{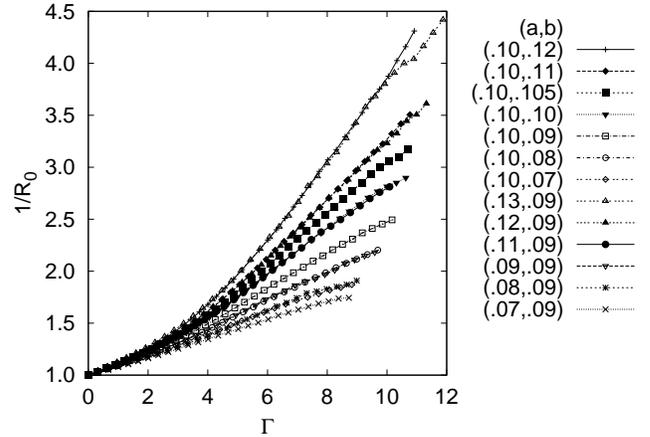}}
\vspace{0.15in}
\caption{Width of the field distribution $1/R_0$ vs the 
$\log$-energy scale $\Gamma$ for the same set of different 
initial conditions as in Fig.~\ref{intercept}. There is a clear
initial transient in all the data for $1/R_0<1.5$ (see also plot 
of $1/R_0$ vs $N$ for our candidate critical point in 
Fig.~\ref{field_distflow}). After the initial transient the data
are consistent with the expected linear behavior at criticality
and with the saturation of the width in the disordered phase.
}
\label{width_vs_gamma}
\vspace{0.15in}
\end{figure}

Since we could not obtain a clear fixed point in the two-parameter
space defined by the simple linear fit to the bond distribution,
we instead chose as our candidates for critical points those 
that produce a scaled bond distribution whose
intercept $P_{\mathrm sc}(0)$ appeared to be stabilizing
to a fixed point value under the action of the RG.  In
running the RG, at each decimation step the maximum
energy term that is ``integrated out'' is either a bond
or a field. The ratio of the frequency of these occurrences 
is simply the
intercept $P_{\mathrm sc}(0)$ (since  we have normalized
 so that the intercept of
the scaled {\it field} distribution is $R_{\mathrm sc}(0)=1$), 
and is easily estimated without fitting any
distributions by counting the number of occurrences of the 
two types of decimations as the RG runs.
Stability of the ratio of the frequencies of the two types 
of decimations is a direct indication of a ``balance''
between the field part of the hamiltonian and the bond part.
A detailed study, extended down to a factor of over $1000$ in $N$,
shows three different types of behavior of the intercept,
which we interpret as follows:
If $P_{\mathrm sc}(0)$ is clearly decreasing towards zero then 
the system is in the disordered phase.  If $P_{\mathrm sc}(0)$
is steadily increasing then the system is in the ordered phase.
Finally, if $P_{\mathrm sc}(0)$ appears to be saturating 
at some value then we have a candidate for the critical point.
Fig.~\ref{intercept} illustrates this.
Starting from different initial two-dimensional lattices and
different initial distributions, we always find that the
apparently ``critical'' (stable) value of the
intercept is in the range $0.1 - 0.2$,
using what we think are conservatively large uncertainties
on when and where the intercept stabilizes. This range of 
candidate critical points, shown by the filled symbols in 
Fig.~\ref{intercept}, are used in all of our error estimates. 

In the simple
approximation to the RG that neglects correlations involving
fields and has $P_0 \equiv P(0) \approx 1/\Gamma$,  
the intercept is $P_{\rm sc}(0) \approx 1/B_0$; 
in general at the critical point $P_0 \approx Z_0/\Gamma$ 
and the intercept is thus $P_{\rm sc}(0) \approx P_0/R_0=Z_0/B_0$.

Fig.~\ref{field_distflow} and Fig.~\ref{bond_distflow} show
the evolution under the RG of the field distribution and the
scaled bond distribution of one candidate for the critical point.
Since the bond distribution is not fully stable, our scaling
analysis of the critical flow, which we discuss next, is
not as certain as our conclusion on the nature of the critical
fixed point: i.e. that it is at infinite randomness.
As we already mentioned, direct scaling with $\Gamma$ requires
estimating the additional parameter $\Omega_0$. 
To estimate the ``tunneling scaling'' exponent $\psi$
we therefore consider the evolution of the width $1/R_0$ of 
the field distribution; this is shown in the inset of 
Fig.~\ref{field_distflow}.
It is expected that $1/R_0 \sim N^{-\psi/d}$ at the critical 
fixed point since at asymptotically low energy scales, for 
which the ``bare'' scale $\Omega_0$ is not important, $1/R_0$ 
should be proportional to $\Gamma$.  
(Fig.~\ref{width_vs_gamma} shows that this is indeed true in 
the later stages of the renormalization after the initial 
transient.)  It can be seen from the inset in 
Fig.~\ref{field_distflow} that during the initial 
transient the width of the field distribution grows more slowly than
later in the renormalization.  This accelerating growth of the
width occurs for all our candidate critical points for $d=2$,
and emphasizes that the RG flow is certainly towards infinite
randomness.  Fitting the later stages of the RG for all the
initial conditions that appear consistent with being
critical -- illustrated by the filled symbols in 
Fig.~\ref{intercept} --  gives exponent estimates in the range 
\cite{stphyspsi}

\begin{eqnarray}
\psi=0.42 \pm 0.06 ~.
\end{eqnarray}
For all candidate critical points the estimated exponent 
$\psi$ is noticeably larger than the $\psi = d/(1+B_0)=0.2-0.3$ 
from the simple approximation discussed above if one uses the 
$1/B_0\cong P_{\rm sc}(0)=0.1-0.2$ obtained from the apparent
intercept; this indicates that correlations between the fields 
and the bonds must be substantial at the critical fixed point.
Indeed, after renormalization, correlations between a field and 
its adjacent bonds are easily detected: the strengths are 
anticorrelated, so that for example
a cluster with a weak renormalized field is more likely to have 
strong renormalized bonds connected to it.

\begin{figure}
\epsfxsize=\columnwidth
\centerline{\epsffile{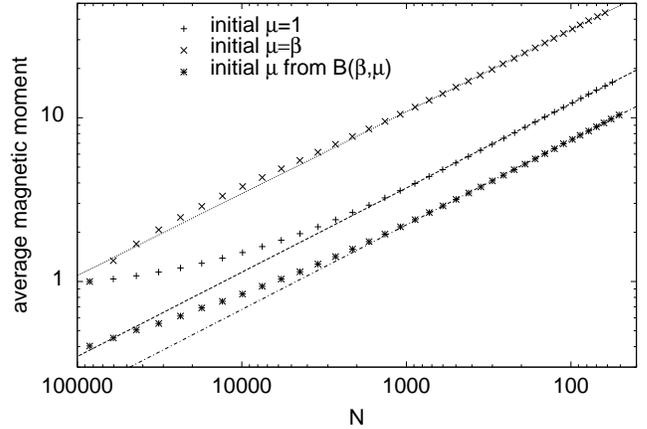}}
\vspace{0.15in}
\caption{Scaling of the average magnetic moment per cluster,
$\overline{\mu}$, with the number of remaining clusters $N$.  
Under the RG, the magnetic moment of a cluster, $\mu_i$, and 
local log-field magnitude, $\beta_i$, become strongly 
positively correlated, and we expect significant
transients if our initial conditions do not have these
correlations.  The initial transient is clearly seen if we
start with $\mu_i=1$ for all sites, or with $\mu_i=\beta_i$ --- 
an attempt to imitate the positive correlation ---, 
but the transient dies off quickly as we run the RG.  
This transient is slightly suppressed if we generate
initial $(\beta_i,\mu_i)$ from the joint distribution function
$B(\beta,\mu)$ which is the fixed point joint distribution in the
uncorrelated-field approximation (discussed in the text) 
with $B(0)=7$.  We consistently find
$\mu \sim N^{-0.50 \pm 0.05}$ implying $d_f=1.0 \pm 0.1$ 
and $\eta =2.0 \pm 0.2$.}
\label{MomentScaling}
\vspace{0.15in}
\end{figure}

The fractal dimension $d_f = \phi\psi$ of the critical clusters
can be obtained directly from the 
RG flows at the critical fixed point.  Fig.~\ref{MomentScaling} 
shows the scaling with $N$ of the average magnetic moment 
(proportional to the number of bare spins) of surviving 
clusters.  Direct fits to such plots for our candidate 
critical points give

\begin{eqnarray}
d_f =1.0 \pm 0.1 ~,
\end{eqnarray}
in contrast to the prediction from the simple approximation 
of $d_f=0.7-0.9~$.  Note, however that if $B_0 \cong 4$ is 
obtained from our $\psi \cong 0.4$ by using the simple 
approximation (but ignoring the estimates of $P_{\rm sc}(0)$), 
the predicted $d_f \cong 1.0$ is close to the value 
obtained from the full RG.  This suggests that the correlations 
between the moment and the field on a cluster are reasonably 
well captured by the simple approximation.

From the scaling relation Eq.(\ref{avcorr}), the average 
critical correlations decay with the exponent 

\begin{eqnarray}
\eta = 2.0 \pm 0.2 ~.
\end{eqnarray}
More direct fits for the exponent $\phi$ alone can
be obtained from plots (not shown)
of the average magnetic moment vs. $R_0$, giving

\begin{eqnarray}
\phi= 2.5 \pm 0.4 ~.
\end{eqnarray}  

The recent quantum Monte Carlo study by Pich {\it et al.} \cite{py}
of the two-dimensional random Ising ferromagnet has found evidence
that the width of the distribution of the logarithms of 
characteristic energies grows with sample size at the quantum 
critical point, as for one-dimension.  
They estimate $\psi \cong 0.4$, in 
good agreement with what we  find from the numerical RG.  They also
measured the spatial correlations 
$G(r) = <\sigma^z_0 \sigma^z_r >$ at
criticality and found that the median (and hence typical) correlation
$G_{\mathrm typ}(r)$
falls off faster than a power of $r$, better  fit by
$-\ln G_{\mathrm typ} \sim r^{\psi_c}$ with $\psi_c \cong 1/3$,
not inconsistent with the scaling prediction $\psi_c=\psi$. 
In contrast, the average critical
correlations exhibit a power-law decay with $\eta \cong 2$, 
which implies that the fractal dimension of the critical cluster is 
$d_f=\phi\psi \cong 1$, again in good agreement with the
exponents estimated from our RG study.

\narrowtext

\section{Ordered and Disordered Phases}

We now turn to a discussion of the ordered and disordered phases. 
Here and henceforth, we will denote the parameter that controls 
the difference between the strengths of the typical random 
fields and those of the typical random bonds in the original 
Hamiltonian by $\delta$, chosen so that the 
zero-temperature quantum critical point corresponds to $\delta=0$, 
the zero-temperature disordered phase to $\delta>0$ and 
the zero-temperature ordered phase to $\delta<0$.

In our numerical RG studies, we do not know the fixed point 
accurately enough and do not have sufficient control over initial
transients to study the off-critical flows directly.  
Nevertheless, we can still obtain some information about the 
near-critical properties indirectly from the critical flows 
(just as, in conventional systems, the correlation length 
exponent, $\nu$, is related to the decay of energy density 
correlations at the critical point).

The effective field of a cluster is generally a product of 
some number $f$ of the {\it original} fields divided by a 
product of $(f-1)$ {\it original} interactions (both the original 
fields and original interactions need not be distinct).  
At the critical point we expect

\begin{equation}
f\sim\Gamma^\rho,
\end{equation}
with the new exponent $\rho$ satisfying

\begin{equation}
\rho\geq \max (\phi, \frac{1}{\psi}).
\end{equation}
The first inequality obtains because any spin which is active 
in a cluster contributes (at least once) its original field 
to the effective field of the cluster; the second inequality 
follows from the observation that the (decimated) bonds which 
hold the cluster together also contribute to the effective 
field and must reach across the diameter of the cluster which 
is of order $\Gamma^{\frac{1}{\psi}}$.  

\narrowtext
\begin{figure}
\epsfxsize=\columnwidth
\centerline{\epsffile{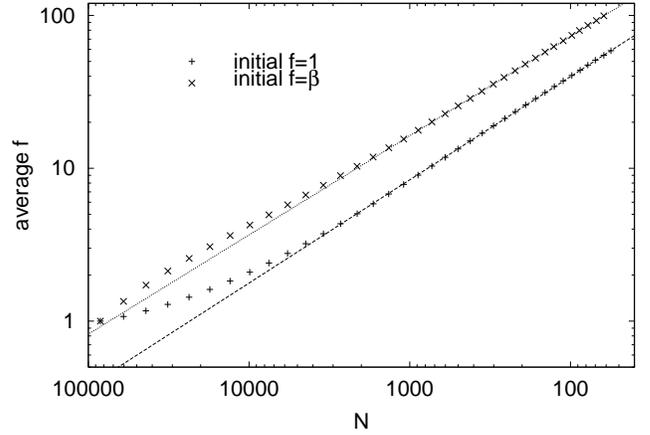}}
\vspace{0.15in}
\caption{Scaling of the average cluster ``history'' $f$---the 
number of the original fields whose product enters the effective 
field---with the number of remaining clusters $N$.  Under the RG, 
$f$ and the log-field magnitude $\beta$ become positively 
correlated, but the transient generating this correlation 
dies off quickly.  By starting either from $f_i=1$
and the corresponding quantity for the bonds $f_{ij}=1$, 
or from $f_i=\beta_i$ and $f_{ij}=\zeta_{ij}$, we consistently
find $f \sim N^{-0.67 \pm 0.07} \sim L^{1.34 \pm 0.14}$.}
\label{HistoryScaling}
\vspace{0.15in}
\end{figure}

An instructive way to understand the effect of deviations from 
criticality, is to move away from it by simply multiplying 
all the original fields by an amount $1+\delta$.  
Perturbatively, this would change the log-fields
at scale $\Gamma$ by of order $f\delta \sim\Gamma^\rho \delta$.
If we neglect the effects of $\delta$ on
changing the order of decimations, we obtain the crossover scale
away from criticality

\begin{equation}
\Gamma_\delta \sim |\delta|^{-\frac{1}{\rho-1}},
\end{equation}
as the scale at which the log-field changes become comparable 
to a typical
log-field or log-interaction ($\sim\Gamma$) and thereby substantially
alter the distribution.  This can be justified as follows: 
Before the scale $\Gamma_\delta$ is reached, $\delta$ 
only changes the global sequence of the decimations but not 
significantly the local sequence; thus the estimate of the 
crossover scale is consistent.  Note also that ``chaotic'' 
behavior under the RG flows, as occurs at (and below) the 
critical point in classical spin glasses,\cite{sgchaos} 
cannot occur here.  This is because the classical Ising model 
that is equivalent to the random quantum Ising model---the higher 
dimensional generalization of the McCoy-Wu model \cite{mw}---is 
purely ferromagnetic; this implies that increasing an original 
field on one spin in the quantum model must increase the 
last-to-be-decimated field in any fixed region around this spin. 

The scaling argument above yields a correlation length
$\xi=\xi_{\mathrm av}\sim|\delta|^{-\nu}$ with

\begin{equation}
\nu=\frac{1}{(\rho-1)\psi}.
\end{equation}
Numerically, we obtain from fitting to $f \sim N^{-\rho\psi/2}$,

\begin{equation}
\rho\psi=1.34\pm 0.14
\end{equation}
(Fig.~\ref{HistoryScaling}) and hence $\nu$. We can also fit 
more directly for $\nu$ using $f R_0 \sim N^{-1/(2\nu)}$, 
obtaining a similar estimate

\begin{equation}
\nu=1.07\pm 0.15,
\end{equation}
which is consistent with the bounds \cite{cc} $\nu \geq 2/d = 1$ and 

\begin{equation}
\nu \leq \frac{1}{{\mathrm max}(1,\phi \psi)-\psi} \cong 1.7.
\end{equation}

\subsection{Disordered Phase: Correlations and Griffiths-McCoy 
Singularities}

In the disordered phase, $\delta>0$, the average spin correlations
will be dominated by rare large clusters and decay
exponentially: 

\begin{equation}
\overline{G(r)} \sim e^{-r/\xi}.
\end{equation}
The typical correlations, on the other hand, will decay more 
rapidly, as 

\begin{equation}
-\ln G_{\mathrm typ}(r)\sim r/\xi_{\mathrm typ}
\end{equation}
with

\begin{equation}\label{xityp}
\xi_{\mathrm typ}\sim\delta^{-(1-\psi)\nu}\sim\xi^{1-\psi}<\xi.
\end{equation}
This can be seen as follows: At the crossover scale away 
from criticality, $\Gamma_\delta$, which corresponds to length 
scale $\xi$, the ratios between typical remaining bonds and 
fields are of magnitude $\ln(J/h) \sim -\Gamma_\delta$.
Two spins of interest separated by distance $r$ much longer 
than $\xi$ will each have the maximum---albeit small---component 
of their spin on a remaining cluster near to them. As these 
clusters and those between them are decimated until eventually 
the two spins are contained in the same (mutual) cluster, 
a multiplicative factor of order the typical $J/h$ ratio at the 
crossover scale will reduce the components of the spins on the 
remaining clusters---and thus on the eventual mutual cluster---for 
each one of the clusters at the crossover scale that is decimated;
i.e., for each element of length of order $\xi$. The result 
Eq.(\ref{xityp}) follows.
Note that the typical correlations at and near criticality  
have the scaling form 

\begin{eqnarray}
-\ln G_{\mathrm typ} \sim r^\psi  F_{\mathrm typ}(r/\xi).
\end{eqnarray}

The behavior of the typical correlations is related to the 
behavior of the distributions of fields and bonds.  The 
field distribution in the disordered phase has finite
width in the limit of low energy (as in one dimension),
$R(\beta;\Gamma |\delta) \cong R_0(\delta) e^{-R_0(\delta)\beta}$
for $\Gamma \gg \Gamma_\delta$, 
while the bonds continue to become weaker. The limiting width is 
of order

\begin{equation}
1/R_0(\delta) \sim \Gamma_\delta \sim \delta^{-\psi\nu}
\end{equation}
which diverges as  $\delta\to 0$.  \cite{dsf,yr}
As in one-dimension, this gives rise to a disordered
``Griffiths'' phase with continuously variable dynamical
exponent $z$ that relates the scales of energy and 
length---the typical fields and spacing of surviving clusters---via 

\begin{equation}
\Omega\sim L^{-z}
\end{equation}
with $z$ diverging as 

\begin{equation}
z\sim\delta^{-\psi\nu}
\end{equation}
for $\delta \to 0$, consistent with the numerical data of 
Ref.~\cite{py}. 

In the disordered phase the distribution of log-fields tends to a 
simple exponential form at low energies with the limiting width 
$1/R_0 \approx z/d$. \cite{yr,dsf}
Concomitantly, there is a continuously variable
power law singularity in the average ground-state magnetization 
per spin in an ordering field $H$:

\begin{equation}
M\sim H^{d/z}|\ln H|^x
\end{equation}
with the exponent of the logarithmic factor not determined 
from these simple arguments.  This gives the leading low-$H$
behavior near the critical point where $z>d$.  For larger 
$\delta$ where $z<d$, this instead gives a singular correction to
an analytic $M(H)$.  The low temperature zero-field 
susceptibility likewise diverges for $z>d$ as $1/T$ raised to a 
continuously varying power that is less than unity --- a weaker 
than Curie law divergence. 

For small (positive) $H$ and small $|\delta|$, the magnetization 
has a scaling form:

\begin{equation} \label{scfncn}
M(H,\delta) \sim |\ln H|^{\phi-d/\psi}\Xi(C\delta|\ln H|^{1/\psi\nu})
\end{equation}
with both $C$ and the cutoff scale implicit in the $\ln H$ 
representing non-universal corrections to scaling. 
When $u$ is large and positive, the scaling function 
$\Xi(u) \sim \exp(-u^{\psi\nu})$ times a power of $u$. 

The clusters surviving to low energies in the disordered phase 
are rare large strongly coupled ferromagnetic clusters which 
exist even in the disordered phase.  The most probable way 
for such a region to occur is (it appears) for there to be, at 
the crossover scale $\Gamma_\delta$, a connected set of 
$n$ clusters each with linear size $\sim\xi$ 
with somewhat anomalously weak fields on them and somewhat 
anomalously strong effective couplings between them; 
these will make them join together at lower energies into the rare 
large cluster of interest.  Note that because the system has 
strong randomness, there is no requirement that this large cluster
has a compact geometry.  Indeed, for large $n$ we expect that 
the most probable such clusters will have, on scales larger 
than $\xi$, a geometry similar to incipient {\it classical}
percolation clusters, as the requirements for a large rare cluster 
to form out of the scale $\xi$ sub-clusters is qualitatively 
like that for a large cluster to form out of the small scale 
objects in conventional percolation.  [See more detailed 
discussion in next sub-section.]  Since the probabilities of the 
occurrence at scale $\xi$ of each of the $n$ such clusters and 
each of the ($\sim n$) corresponding couplings are 
determined primarily by local properties and are hence roughly 
independent, the probability of the rare large cluster is of order 
$1/\alpha^n$ with $\alpha>1$; $\alpha$ is independent of 
$\delta$ for small $\delta$, since the needed clusters and 
bonds are not very atypical at scale $\xi$.  The field on such 
a cluster will be of order the energy scale for crossover away 
from criticality, multiplied by of order $n$ factors, each less 
than unity with logarithm of order $\Gamma_\delta$:

\begin{equation}
\tilde{h} \sim \Omega_0 (e^{-a\Gamma_\delta})^n
\end{equation}
with $a$ independent of $\delta$ near the critical point.  
The probability of a cluster with an anomalously small 
field $\tilde{h}$ is thus approximately 

\begin{equation}
p(\tilde{h})\sim e^{-(\ln\alpha)\ln(\Omega_0/\tilde h)
/a\Gamma_\delta}.
\end{equation}
If one finds the least rare such clusters, i.e. with $(\ln\alpha)/a$ 
as small as possible, then 
$R_0(\delta)=(\ln \alpha)/a \Gamma_\delta$ and these rare 
clusters give rise to the singular scaling in the Griffiths 
phase quoted above. As stated earlier the dynamic exponent $z$ is 
given by the asymptotic low energy value of $d/R_0$. [Note that we 
have ignored here differences 
between $\ln\Omega_0$ and 
$\ln\Omega_\delta = \ln\Omega_0-\Gamma_\delta$ 
since these will not affect the dominant behavior in the limit 
of interest.  Also note that since $|\ln \tilde{h}| \sim n$,
a factor $p(\tilde{h})$ controls the asymptotic behavior of the 
distribution of the $|\ln \tilde{h}|$-variable.]

The RG generates the low-energy tail $\sim e^{-R_0(\delta)\beta}$ 
of the log-field distribution on scales $L \gg \xi$ precisely 
because the rare large clusters discussed above are not 
{\it too} rare.
Note that an exponential tail of the distribution of the cluster 
effective fields is generated by the RG almost immediately and 
some rare arbitrarily large clusters are present at any energy 
scale, as long as there is no infinite cluster that would 
consume them.  In particular, this tail is also generated at 
the critical point, but it continues to become longer as the 
energy scale decreases.

\subsection{Ordered Phase: Percolation and Finite Temperature 
Ordering}

The behavior in the ordered phase differs strikingly from
that in one dimension.  At a {\it finite} energy scale

\begin{equation}
\Omega_\infty=\Omega_0 e^{-\Gamma_\infty}\sim 
\Omega_0 \exp[-K|\delta|^{-\psi\nu}],
\end{equation}
a single infinite cluster (with zero transverse field) develops.
Finite clusters, some of which will join the infinite cluster
at lower energies, coexist with it.  The spontaneous magnetization 
is proportional to the number of sites in the infinite cluster at
$\Omega\to 0$ yielding 

\begin{equation}
M_0\sim|\delta|^\beta
\end{equation}
with

\begin{equation}
\beta=\nu\eta/2=\nu(d-\phi\psi).
\end{equation}
This implies that the scaling 
function in Eq.(\ref{scfncn}) has the asymptotic form in the
ordered phase 
$\Xi(u \rightarrow -\infty) \sim (-u)^{d\nu -\phi\psi\nu}$. 

As a consequence of the infinite cluster development, 
in the ferromagnetically ordered phase there is, 
in contrast to one-dimension, 
an interfacial energy density---albeit exponentially 
small---for a domain wall for any $\delta<0$.  This 
`surface tension' scales as 

\begin{equation}
\sigma\sim\xi^{1-d}\Omega_\infty
\end{equation}
---simply the energy needed to cut the infinite cluster in half.  
The transition temperature for destruction of the long range order
by thermal fluctuations will similarly be determined by the
strength of the bonds that hold together the infinite cluster; thus 

\begin{equation}
T_c\sim\Omega_\infty.  
\end{equation}
We note that the arguments given below imply that the width 
of the classical finite temperature critical region as the 
temperature is reduced at fixed negative $\delta$, will be 
of order $T_c$.

The behavior near the energy scale $\Omega_\infty(\delta)$ at 
which the infinite cluster forms has aspects similar to 
conventional classical percolation (in contrast to the 
zero-temperature quantum percolation
transition at $\delta=0$).  At the 
crossover scale $\Gamma_\delta = \ln(\Omega_0/\Omega_\delta)$ 
away from criticality, the bonds will start to dominate over 
the random fields.  At this scale the sizes of the clusters
and the lengths of the  
bonds will typically be of order the crossover 
length scale $\xi \sim \Gamma_\delta^{1/\psi}$; indeed, bonds 
much longer than this will be exponentially rare.  Between the 
log-energy scales $\Gamma_\delta$ and 
$\Gamma_\infty \approx \ln (\Omega_0/\Omega_\infty)$, at 
which the infinite cluster forms, most of the decimations 
will be of bonds, resulting in the joining together
of clusters.  The process 
of decimation of bonds (and occasional clusters) will continue 
with larger and larger clusters forming until the percolation 
scale $\Gamma_\infty$.  Note that $\Gamma_\infty$ 
will be a fixed (order-one) multiple of the somewhat arbitrarily 
defined crossover scale $\Gamma_\delta$, but the corresponding
physical energy $\Omega_\infty$ is actually exponentially smaller
than the energy $\Omega_\delta$ (for large $\Gamma_\delta$).

Unlike the case at the quantum critical point, the process by 
which the large clusters are joined together as the energy
scale is decreased near the percolation scale is basically 
{\it local}.  The key feature of this locality, which occurs when 
$\Gamma$ is in the range $\Gamma_\delta < \Gamma < \Gamma_\infty$, 
is that  when $\Gamma$ is changed by a small amount, whether, say, 
a large cluster $A$ 
will become joined to a large cluster $B$, and whether the same 
cluster $A$ will become joined to another large cluster $C$, 
are roughly independent events, each only depending on the smaller 
clusters and bonds -- which have typical length scale $\xi$ 
and log-energy scale $\Gamma_\delta$ -- in the vicinity of 
the respective potential connections.  As the percolation scale 
is approached, we expect that this independence will become more 
and more pronounced as the important connections that make the 
large clusters grow become further and further apart.  On the 
basis of this argument, we conjecture that the percolation that 
occurs in the cluster RG at scale $\Gamma_\infty$ is in the 
{\it universality class of classical percolation}, with 
$(\Gamma_\infty - \Gamma)$ playing the 
role of $(p_c - p)$ in classical percolation. 
[Note that the width of the log-bond
distribution at scale $\Gamma_\delta$ is of order $\Gamma_\delta$ 
and will remain so at all scales 
$\Gamma_\delta < \Gamma < \Gamma_\infty$].

The nature of the percolation process at the scale $\Gamma_\infty$ 
controls the critical behavior associated with the 
finite-temperature ordering transition at $T_c \sim \Omega_\infty$ 
over a substantial region of the $T - \delta$ plane. 
On the logarithmic temperature scale, $\Gamma_T = \ln (\Omega_0/T)$,
the finite-temperature spin-spin correlation length, $\xi_T$,
is simply the characteristic length scale of the distribution of 
cluster diameters at scale $\Gamma_T$: clusters that exist at this 
scale will each consist of sets of well-correlated (active) spins, 
while the correlations between these clusters will be destroyed 
by the thermal fluctuations.  On temperature scales above 
$\Omega_\delta$ ($\Gamma_T<\Gamma_\delta$), $\xi_T$ will be given 
by the quantum critical length scale at log-energy scale 
$\Gamma_T$.  But on scales between $\Omega_\delta$ and 
$\Omega_\infty$ ($\Gamma_\delta < \Gamma_T < \Gamma_\infty$), 
the percolation process will cause the correlation length to 
diverge as 

\begin{eqnarray}
\label{xip}
\xi_T \sim \xi 
\left(\frac{\Gamma_\delta}{\Gamma_\infty-\Gamma_T}
\right)^{\nu_p} ~,
\end{eqnarray}
where $\nu_p$ is the {\it classical} percolation correlation 
length exponent --- with $\nu_p=4/3$ in two dimensions --- and 
$\xi\sim (-\delta)^{-\nu}$ is the correlation length associated 
with the {\it quantum} critical point (i.e., the characteristic 
length at the crossover scale $\Gamma_\delta$).

As the critical temperature is approached, the RG approximation 
will eventually break down at any non-zero $\delta$. The 
clean separation of bonds into `strong' for those stronger 
than $\Gamma_T$,  and `weak' for  those weaker than $\Gamma_T$ 
will not hold for the `marginal' bonds whose strength is of 
order $T$, which correspond to those whose log-strength is 
$\Gamma_T \pm O(1)$.  This implies an $O(1)$ multiplicative 
uncertainty in the proportionality between $T_c$ and $\Omega_\infty$
and it also implies that the percolation-dominated form of 
the critical behavior of Eq.(\ref{xip}) breaks down when 
$\Gamma_\infty-\Gamma_T $ is of order one.  Closer to the 
finite-temperature critical point, the behavior will be dominated 
by the thermal fluctuations of the marginal bonds that 
link very large almost percolating clusters.  This will make 
the critical behavior cross over to that of the conventional 
{\it classical d-dimensional random bond Ising} universality 
class with 

\begin{eqnarray}
\xi_T 
\sim \xi \Gamma_\delta^{\nu_p} \left(\frac{T_c}{T-T_c}\right)^{\nu_I}
\sim \xi \Gamma_\delta^{\nu_p}
\left(\frac{1}{\Gamma_\infty-\Gamma_T}\right)^{\nu_I}~,
\end{eqnarray}
where $\nu_I$ is the classical random Ising correlation length 
exponent, equal to one in two dimensions (assuming the interactions
are not frustrated).

As a function of temperature, this double crossover in the 
critical behavior will be particularly hard to observe due to 
the logarithmic temperature scale which makes the crossover 
energy scale $\Omega_\delta$ exponentially larger than $T_c$ 
for small (negative) $\delta$.  But if the temperature is held 
fixed and extremely small---i.e. $\Gamma_T\gg 1 $--- then the 
crossovers can be seen more readily by decreasing the relative 
strengths of the random fields that are parametrized by $\delta$.
The critical value, $\delta_c$, is of order 
$-\Gamma_T^{-1 / \psi\nu}$.  As the random fields are reduced 
from $\delta$ of order one until 
$\delta \sim +\Gamma_T^{-1 / \psi\nu}\sim |\delta_c|$, 
the thermal effects will be negligible and $\xi_T$ will diverge 
with the zero-temperature {\it quantum} exponent that we have 
denoted simply $\nu$.  As $\delta$ is further decreased through 
zero until $\delta-\delta_c \sim \Gamma_T^{-1-1 / \psi\nu} 
\sim |\delta_c|/\Gamma_T$, the intermediate classical-percolation 
dominated critical behavior as in Eq. (\ref{xip}) will obtain with 
the exponent $\nu_p$.  Finally, for 
$\delta-\delta_c \ll \Gamma_T^{-1-1 / \psi\nu}$ the 
classical random Ising critical exponent $\nu_I$ will control 
the divergence of $\xi_T$.  Note that in the limit of 
asymptotically small $1/\Gamma_T$, all three of these regimes 
will become very broad on a $\ln ((\delta_c -\delta)/ \delta_c)$ 
plot.

\subsection{Ordered Phase: Singularities}
 
We now turn to properties of the ordered phase on energy and 
temperature scales much lower than the ordering temperature $T_c$.
The low-energy properties of the ordered phase 
for $d > 1$ will {\it not} have the strong power-law Griffiths
singularities found in the disordered phase and the
one-dimensional ordered phase.  This can be seen
in the RG language which naturally incorporates the role of rare 
anomalous regions.  If we continue the RG much below 
$\Omega_\infty$, i.e., beyond the formation of the infinite 
cluster, we will find that the remaining finite clusters are 
almost always  connected only to the macroscopic cluster and at 
lower energy scales almost all of these will
either be decimated or will join the infinite cluster; 
fewer and fewer will join together to make 
larger finite clusters.  Since the finite clusters and the 
bonds connecting them to the macro-cluster are decimated 
independently, no new clusters or bonds will be generated, and the 
low-energy tails of the distributions remain essentially the same 
as they were just below $\Omega_\infty$ (when the widths of 
{\it both} the log-field and the log-bond  distributions are 
of order 
$\sim \ln(\Omega_0/\Omega_\infty) \sim \Gamma_\delta 
\sim |\delta|^{-\psi\nu}$).  

As in the disordered phase, the pre-formed tails of the field 
and bond distributions represent rare large regions responsible 
for the low-energy excitations in the system.  The role of 
these rare fluctuations, however, is very different in the 
ordered phase; although they still make the system gapless, 
they do not dominate all the low-energy properties (such as, 
for example, the response to a small ordering magnetic field). 
In contrast to the disordered phase, the dominant rare regions 
in the ordered phase are indeed {\it very rare} and
do not produce a power-law singularity in the density of states at 
zero energy.  This can be seen by analyzing the probability 
that a cluster with a very small effective field $\tilde{h}$
survives down to energy scale 
$\Omega \sim \tilde h \ll \Omega_\infty$.  
We can consider such a surviving ferromagnetic cluster 
to be composed of $n$ subclusters (each of diameter $\sim \xi$) 
with $\tilde h \sim \Omega_0 e^{-cn\Gamma_\delta}$, 
as in the disordered phase.  But this cluster must be isolated 
very effectively from the rest of the system---with effective 
coupling linking it to the infinite cluster of order $\tilde h$ 
or weaker.  In the disordered phase, the typical length of a bond 
with effective coupling 
$|\ln \tilde{J}| \sim n\Gamma_\delta$ is $L \sim n\xi$. 
Thus to achieve sufficient isolation, the disordered region 
around the droplet must have a {\it linear} size of order 
$\sim n\xi$.  The probability of such a rare region is very 
small---of order $1/\alpha^{n^d}$---so that the generic 
low-energy tail that the RG can generate is 

\begin{equation}
R(|\ln \tilde{h} |) \sim e^{-\tilde c |\ln \tilde h|^d},
\end{equation}
very different from the $\sim e^{-\tilde c |\ln \tilde h|}$ 
tail in the disordered phase and the similar tail in the 
one-dimensional ordered phase.  It is also strikingly different 
from an even longer tail,  
$\sim e^{-\tilde c |\ln \tilde h|^{1-1/d}}$, that occurs in the 
ordered phase of the {\it dilute} quantum Ising system of 
Ref.~\cite{ss} which differs from ours by some fraction of the 
initial $J_{ij}$'s being {\it zero}, so perfectly isolated 
clusters can form even in the ordered phase. 
The origin of the difference between these cases is easy 
to understand: in one-dimension, the length and the volume 
of an isolating region are the same, while in the dilute case, with 
a delta-function weight at zero coupling, a droplet of size $n$ can 
be {\it completely} isolated by a  ``disordered'' region of 
volume $n^{(d-1)/d}$---just a surrounding surface of missing bonds.

\section{Higher Dimensions and Related Systems}

\subsection{Higher Dimensional Random Ferromagnetic 
Quantum Ising Models}

So far, we have presented detailed results only for two dimensions, 
although the general scaling picture, exponent equalities, 
behavior of correlation functions, etc., should be qualitatively 
the same for strong randomness in any dimension ($d>1$) 
for which there is a stable infinite-randomness critical fixed point.
Our numerical studies in three dimensions are sufficient to 
indicate that the infinite-randomness fixed point is stable, 
although they are not thorough enough to yield reliable estimates of 
exponents and their uncertainties.  
Since in both two and three dimensions, weak 
randomness is a relevant perturbation away from the pure 
fixed point, we expect the same strong-randomness-dominated
critical behavior to occur even for arbitrarily weak randomness. 

The situation in higher dimensions -- $d\geq 4$ -- is far more 
uncertain.  It is not clear at this point whether or not the 
direction of the RG flow at strong randomness reverses 
for $d$ sufficiently large.

For weak disorder, it would appear that the situation is 
clearer: the Harris criterion would seem to indicate that 
for $d>4$ weak randomness is irrelevant.  But one must be 
very careful.  There are other situations known in which 
weak randomness formally appears to be irrelevant but for 
which exponentially rare regions change the behavior for 
arbitrarily weak randomness. \cite{dsfcdw}
We strongly suspect, as argued below, that this will be 
the case here.
In general, Griffiths singularities and other strong-randomness-like 
effects will start to appear when the random quantum Ising 
system is close enough to the quantum transition that the 
distribution of $J$'s and the distribution of $h$'s overlap 
in the sense that for some values of $(h,J)$ in the support 
of these distributions, a {\it pure} system would be in the 
ordered phase, while for other values of $(h,J)$, a pure 
system would be in the disordered phase.  This implies that 
arbitrarily large rare regions will exist that act as if 
they were in the opposite phase than the full system is.  
In particular, in the disordered phase sufficiently close 
to the critical point, strongly correlated 
clusters will exist with broadly distributed effective fields 
and effective interactions between them which are broadly 
distributed and typically decay exponentially with their 
separation.  As the quantum critical point is approached, 
these rare clusters and their couplings will effectively act 
like a strongly random system which we expect will dominate 
the behavior and cause the whole system to be driven to strong 
(but not necessarily infinite)
randomness sufficiently close to the critical point -- however 
weak the original randomness.  This 
intriguing possibility clearly merits further investigation. 

\subsection{Other Quantum Transitions with Discrete Broken 
Symmetries}

As mentioned in the Introduction, the infinite randomness 
critical fixed points found here control more than just 
Ising ferromagnetic quantum transitions.
In particular, as pointed out for the one-dimensional case 
by Senthil and Majumdar \cite{pot}, Potts models or any random 
quantum systems with continuous (second order) transitions 
at which a discrete symmetry of a non-conserved order parameter
is broken, will have the same 
critical behavior as the Ising case, with the extra degrees 
of freedom just ``going along for the ride'' on the basic 
geometrical transition.

This holds even for systems which are frustrated on small 
length scales, such as quantum Ising spin glasses.  Because 
of flow to the infinite randomness fixed point, the 
frustration will become irrelevant at low energies at the  
critical point, since in any frustrated loop the weakest 
interaction will be infinitely weaker than the others, so can 
be ignored.  
The primary changes here concern the coupling to a uniform 
magnetic field in the $z$-direction, and the behavior at nonzero
temperature in the ordered phase.  Because the uniform field 
is not an ordering field for the spin glass, the magnetic moment 
of the large clusters will be random in sign, scaling as the 
square root of the number of active spins on the cluster.  
At the critical point this will change the $M(H)$ scaling, 
yielding

\begin{equation}
M_{SG} \sim |\ln H|^{\frac{1}{2}\phi-d/\psi}
\end{equation}
in contrast to Eq.(\ref{McH}) for the ferromagnetic case.
In the disordered phase, $M$ will scale as the {\it same} 
power of $H$ as in the ferromagnetic case with only the 
logarithmic prefactors modified.  In the ordered phase, 
the behavior of the nonzero-temperature, long-distance 
correlations will cross over to classical spin glass
behavior at and below a temperature of order $\Omega_{\infty}$.
For $d=2$, true long-range spin-glass order will be present only
at zero temperature, because the lower critical dimension for
the classical spin glass is always more than two.

\subsection{Random Quantum XY And Heisenberg Antiferromagnets}

The simplest example of an infinite randomness quantum 
fixed point occurs for one-dimensional random Heisenberg 
(or XY) spin chains.  In the corresponding phase, 
the ``random singlet phase'', each spin is paired in a 
singlet with one other spin, usually one close by, but a 
small fraction of the spins are paired very weakly with 
spins far away.  The RG analysis of this system, first 
carried out by Ma, Dasgupta and Hu \cite{mdh} and then 
more fully by one of us \cite{f94}, is a simpler version 
of that used in the present paper. 

A similar RG analysis was carried out for two- and 
three-dimensional random antiferromagnets by Bhatt and Lee 
\cite{bl} over a substantial range of energy scales, 
in particular including those relevant for experiments 
on the insulating phase of phosphorus-doped silicon.
This investigation has been extended by two of us\cite{dahrs} 
to the strong-randomness limit.  We have found that for 
$d \geq 2$, in contrast to one-dimension, the 
infinite-randomness random-singlet fixed point of random
Heisenberg or XY quantum antiferromagnets is {\it unstable} 
towards a state with finite randomness and, presumably, 
more conventional scaling; this state includes both 
antiferromagnetic and ferromagnetic effective interactions, 
and involves clusters with moments much larger than those of 
the single spins that dominate the low energy behavior in 
the one-dimensional case.

\section{Summary}

In summary, we have studied  random quantum Ising ferromagnets 
using an energy space cluster RG which becomes exact for 
strong randomness.  Using the structure of the RG,
we presented a scaling picture of the behavior near 
an infinite-randomness quantum critical fixed point that 
can occur.  Near to this fixed point -- corresponding to low 
energy scales near the zero-temperature quantum phase 
transition -- the RG yields asymptotically exact results. 
We have implemented the RG numerically, primarily 
in two dimensions, and found that the critical behavior is 
indeed controlled by such an infinite-randomness
fixed point, as in one-dimension. We estimated 
numerically the corresponding critical exponents in 
two-dimensions, and discussed the properties of the 
disordered and ordered phases.  In the
disordered phase we found that rare anomalously strongly coupled 
ferromagnetic clusters---in the RG language, a low-energy tail 
of the cluster field distribution generated by the decimation 
procedure---dominate the low-energy behavior and cause  
power-law Griffiths-McCoy singularities near 
the phase transition.  In the ordered phase for $d > 1$, on the
other hand, the Griffiths singularities are much weaker,
and do not produce divergences in thermodynamic quantities;
the low-energy density of states they produce vanishes 
faster than any power of the energy.

The universality class controlled by the infinite randomness 
quantum Ising critical fixed point is very broad; it includes 
all continuous quantum transitions in random systems at which 
a discrete symmetry is broken; since in two dimensions, first 
order transitions are not possible in random systems, this 
class should include {\it all discrete-symmetry-breaking 
transitions} providing there are no conservation laws which 
alter the quantum dynamics in an essential way (for an example 
of an Ising case with a conserved order parameter, see \cite{f94}).
The nature of the discrete symmetry breaking quantum 
transitions we have studied is controlled by a 
novel type of percolation --- rather surprising given the 
intrinsic quantum nature of the underlying models.  As the 
rules of this percolation process are asymptotically 
{\it classical} (although the process is {\it not} conventional 
percolation), one might hope that conformal field theory 
approaches which take advantage of the two-dimensional 
(rather than ``$2+1$'' dimensional) structure could perhaps 
be used to obtain analytic results for some of the properties 
of such two-dimensional random quantum systems.

We thank Ravin Bhatt, Kedar Damle, Matthew Hastings and Peter Young 
for helpful discussions. 
Support for this work was provided by the
National Science Foundation through grants
DMR-9400362, -9630064 and -9802468 and via Harvard's MRSEC.



\begin{references}
\bibitem{stphyspsi}  Preliminary results quoted in \cite{stph} 
had not gotten past the initial RG transient and suggested a 
smaller $\psi$ of order $0.2$.
\end{references}
\end{document}